\documentclass[12pt]{article}

\newcommand{\be}{\begin{equation}}
\newcommand{\ee}{\end{equation}}
\usepackage{amsmath}
\usepackage{amssymb}
\usepackage{latexsym}
\usepackage{epsfig}

\newcommand{\Z}{\mathbb{Z}}
\newcommand{\R}{\mathbb{R}}

\newcommand{\bea}{\begin{eqnarray}}
\newcommand{\eea}{\end{eqnarray}}

\newcommand{\ti}{\tilde}

\newcommand{\RRR}{{\hbox{\rm R\kern-2.35mm R}}}

\def\ZZZ{{\hbox{ Z\kern-1.6mm Z}}}

\newcommand{\sectiono}[1]{\section{#1}\setcounter{equation}{0}}

\def\a{\alpha}
\def\b{\beta}

\def\d{\delta}
\def\g{\gamma}

\def\l{\lambda}

\def\r{\rho}                                     
\def\s{\sigma}                                   

\def\z{\zeta}
\def\D{\Delta}

\def\L{\Lambda}

\begin{document}

\begin{titlepage}
\rightline{June 2014}
\rightline{  Imperial-TP-2014-CH-02}
\begin{center}
\vskip 2.5cm
{\huge \bf {
Finite   Gauge Transformations and Geometry in Double Field Theory}}\\
\vskip 2.0cm
{\Large {  C.M.Hull  }}
\vskip 0.5cm
{\it {The Blackett Laboratory}}\\
{\it {Imperial College London}}\\
{\it {Prince Consort Road}}\\
{\it { London SW7 @AZ, U.K.}}\\

\vskip 2.5cm
{\bf Abstract}
\end{center}

\vskip 0.5cm

\noindent
\begin{narrower}
Recently proposed forms for gauge transformations with finite parameters  in double field theory are discussed and problematic issues are identified. A new form for   finite gauge transformations is derived that reveals  the underlying gerbe structure and the close relationship with generalised geometry. 
The nature of generalised tensors is elucidated, and in particular it is seen that the presence of a constant metric with split signature    does not restrict the doubled geometry, provided it is a   generalised tensor rather than a conventional tensor.

\end{narrower}

\end{titlepage}

\newpage

\tableofcontents
\baselineskip=16pt
 
\section{Introduction}

For string theory on a product $\R ^{n-1,1}\times T^d$ of a $d$-torus and Minkowski space, it has been known since the early days of string theory that the $d$ periodic coordinates $x^i$ on the torus are supplemented by 
$d$ dual periodic coordinates $\ti x_i$ conjugate to the  winding numbers, and that the interactions depend on both $x$ and $\ti x$.  The string theory has an $O(d,d;\Z)$ T-duality symmetry acting linearly on the $2d$ coordinates $(x^i,\ti x_i)$. The  $O(d,d;\Z)$ then acts geometrically on the doubled torus $T^{2d}$ with coordinates
$(x^i,\ti x_i)$ through large diffeomorphisms preserving the metric $ds^2=2 dx^id\ti x_i$ with signature $(d,d)$.
In
\cite{Giveon:1991jj}, it was shown that this extends to curved  backgrounds with a $T^d$ torus fibration, with an $O(d,d;\Z)$ T-duality symmetry  provided all fields are independent of the 
torus coordinates. A T-duality invariant string field theory for strings on $\R ^{n-1,1}\times T^d$ was constructed in \cite{Kugo:1992md}. In \cite{Tseytlin:1990va}, string theory on  a toroidal background was argued to lead to an effective field theory on a doubled space, and generalisations to strings with chiral WZW interactions were considered.

In \cite{Hellerman:2002ax,Dabholkar:2002sy,Kachru:2002sk}, it was found that T-duality allows the construction of certain  non-geometric backgrounds.
In \cite{Hull:2004in}, T-folds were introduced as a class of non-geometric backgrounds that include the examples of \cite{Hellerman:2002ax,Dabholkar:2002sy,Kachru:2002sk} and which
  look like manifolds with smooth tensor fields locally but have T-duality transition functions. More precisely, they can be covered with patches of the form $U\times T^d$ where $U$ is a patch of $\R ^n$ and the transition functions involve diffeomorphisms, antisymmetric tensor gauge transformations and $O(d,d;\Z)$ T-duality transformations \cite{Hull:2004in}.
(Generalisations to U-folds with U-duality transitions and mirror-folds with mirror symmetry transitions were also introduced in \cite{Hull:2004in}.) 
Conventional formulations of string theory (e.g.using non-linear sigma-models) cannot be used for such non-geometric backgrounds.
In \cite{Hull:2004in}, it was shown that T-folds can be formulated in terms of  a smooth 
doubled geometry. Replacing the torus fibres $T^d$ with doubled tori so that the patches become
$U\times T^{2d}$, the T-fold transition functions lead to a construction of a smooth manifold with a $T^{2d}$   
fibration, the key point being that the T-duality transitions now act geometrically on the doubled torus fibres as large diffeomorphisms.
This smooth doubled geometry allowed a formulation of string theory on a  T-fold as a constrained  sigma model  with the doubled geometry as the target space \cite{Hull:2004in,Hull:2006va}.

In \cite{Shelton:2005cf,Dabholkar:2005ve}, it was found that there are yet further non-geometric backgrounds that are not even geometric locally -- i.e. they are not constructed from geometric patches. It was proposed in \cite{Dabholkar:2005ve} that many of these are backgrounds in which fields have non-trivial dependence on dual coordinates $\ti x$, and this was verified at special points in the moduli space at which the background reduced to an asymmetric orbifold \cite{Dabholkar:2005ve}. Such doubled geometries were explored further in \cite{Hull:2005hk, Hull:2007jy,Hull:2009sg}.

 In \cite{Dabholkar:2005ve}, it was proposed that the natural 
framework for formulating string theory for such non-geometric backgrounds would be in terms of a string field theory similar to  that of \cite{Kugo:1992md}, and would lead to a theory of  dynamical fields on the doubled geometry.
Such a
Double Field Theory (DFT)  for toroidal backgrounds $\R ^{n-1,1}\times T^d$
was constructed (to cubic order in fields)  in 
\cite{Hull:2009mi}, 
where it was derived from closed string field theory.
It gives
 a theory of fields on the doubled space including  $g_{mn}(x,\ti x)$,   $b_{mn}(x,\ti x)$ and   $\phi(x,\ti x)$.
 It is notationally convenient to supplement the coordinates $y^\mu$ of $\R ^{n-1,1}$ 
 with dual coordinates $\ti y_\mu$, and require all fields to be independent of $\ti y_\mu$ (corresponding to the absence of winding in the non-compact dimensions). Then the space-time $\R ^{n-1,1}\times T^d$
 has $D=n+d$ coordinates $x^m=(y^\mu,x^i)$ and there is a doubled space 
 $\R ^{2n-2,2}\times T^{2d}$ with coordinates $X^M=(x^m,\tilde x_m)$ with $M=1,\dots 2D$, where the dual coordinates are $\ti x_m= (\ti y_\mu, \ti x_i)$.
The constant metric
 $\eta_{MN}$ of signature $(D,D)$ given by
 $$
 ds^2 = \eta_{MN} dX^M dX^N = 2 dx^md \ti x_m$$
  is used to raise and lower indices.
 The $L_0-\bar L_0=0$ constraint of string theory imposes that all fields and parameters $A$ (with matched levels $N=\bar N$)
 satisfy the weak constraint
 \be\label{weak}  
  \partial^{M}\partial_{M} A \ \equiv \  
  \eta^{MN}\partial_{M}\partial_{N} A \ = \ 0
    \ee
(More generally, fields arising in string theory at levels $N,\bar N$ would satisfy $  \partial^{M}\partial_{M} A =N-\bar N$.)
This theory was constructed to cubic order in the fields in  \cite{Hull:2009mi}, and is expected to be non-local at higher orders. In this theory, there is non-trivial dynamics in all $2D$ dimensions, so that the extra dimensions are  truly physical.

The full double field theory with dynamical double geometry has so far proved rather intractable.
A much simpler sub-sector of DFT is obtained by imposing 
 the strong constraint that $ \partial^{M}\partial_{M}=0$
when acting on all fields  and   their products, so that $\partial^M\partial_M A = 0$  
 and $\partial^M A\, \partial_M B=0$
for any fields or gauge  parameters $A$ and $B$. This drastically truncates the theory 
to one that can be constructed to all orders in the fields  \cite{Hull:2009zb,Hohm:2010jy,Hohm:2010pp}.
The (strongly constrained) DFT is a field theory on the doubled space $ M$ with a rich symmetry structure. 
In the remainder of this paper, we will address only DFT with the strong  constraint.
There is now an extensive literature on the subject; see  \cite{Aldazabal:2013sca,Berman:2013eva, Hohm:2013bwa} for recent reviews of DFT with the strong constraint and further references.

The truncation to the strongly constrained theory
typically  results  in fields depending only on half the coordinates, the $x^m$ say,
leading to a conventional field theory on the space parameterised by the $x^m$.
The formulation of strongly constrained DFT of  \cite{Hohm:2010jy,Hohm:2010pp} is background independent (in the sense that it does not depend on a background generalised metric) and has the possibility of  being formulated on more general doubled manifolds $M$ than the product of a torus and flat space for which it was derived.
If the fields have support only on a $D$-dimensional submanifold $N\subset   M$, so that the fields depend only on the coordinates $x^m$ of $N$ and are independent of the remaining coordinates $\ti x_m$, then
 the double field theory essentially recovers Siegel's duality-covariant formulation of gravity and supergravity theories  on $N$ \cite{Siegel:1993th}. 
 This is a conventional field theory on the space-time $N$, 
with massless fields that include a metric $g_{mn}$, a $b$-field $b_{mn}$ and dilaton $\phi$. 
The symmetries of the theory include the diffeomorphisms of $M$ and the antisymmetric tensor  gauge transformations, giving the gauge symmetry group ${\rm Diff} (N)\ltimes \Lambda ^2_{closed} (N)$.
However, the   formulation arising  has a manifest T-duality symmetry.
On $M=\R^{2D}$, the theory has $O(D,D)$ symmetry,
while on  a product $\R^{2n}\times T^{2d} $ of flat space and a torus this is
 broken to a group containing
$O(n,n)\times O(d,d;\Z)$.
This duality-covariant formulation is closely related to the formulation of gravity and supergravity theories in terms of generalised geometry, as in \cite{Grana:2008yw,Coimbra:2011nw}.

 More generally, this picture need only be true locally: in each coordinate patch ${\cal U}$  of the doubled space, there are preferred coordinates $X^M$  and a constant metric $\eta_{MN}$, and the fields
 satisfy the strong constraint. The constraint implies that
  the fields depend on only half the coordinates and so are fields on a $D$-dimensional sub-patch $U\subset {\cal U} $,  where  $U$ is a  subspace of $\cal U$ that is totally  null with respect to $\eta$.
  In general these patches  $U $ need not fit together to form a $D$-dimensional submanifold, but instead  can constitute patches of a non-geometric space such as a T-fold \cite{Hull:2004in}. In each patch, there is a conventional (super)gravity field theory, 
formulated in a duality symmetric way,
but in the non-geometric case they need not fit together to give (super)gravity field theory on a conventional space-time.  Finding a formulation of the theory in such backgrounds  was one of the motivations for seeking a double field theory \cite{Dabholkar:2005ve,Hull:2009mi}.

The background independent formulation of  \cite{Hohm:2010jy,Hohm:2010pp}  gives a  DFT 
on the patch ${\cal U} $ for arbitrary fields $g_{ij} ( x), b_{ij}(x),\phi(x)$ on $U $. An important issue is what transition functions are used to glue these patches together, and what kinds of doubled space $M$ can  arise.
If the constant matrix $\eta$ were to be viewed as a metric tensor on  $ M$, then the presence of a flat metric on
 $ M$ would be  highly constraining, so that  $ M$  would be  locally flat. 
However, it is natural to use the symmetries of DFT in the transition functions -- as usual, patching with symmetries of the theory should lead to well-defined physics. 
If  the matrix $\eta $ is  not a tensor on  $ M$ but is a \lq generalised tensor' transforming with the generalised Lie derivative arising in the DFT gauge transformations, then it can be extended to the whole manifold, apparently without further constraining $M$. This is because the constant $\eta$ is invariant under the DFT symmetries  \cite{Hohm:2010jy,Hohm:2010pp} and so patches smoothly using DFT transition functions. 
To 
 explore this idea and its consequences further, it is necessary to
 understand the geometry of  generalised tensors better.
The constraint 
$ \eta^{MN}\partial_{M}\partial_{N} A  = 0$ would usually require that $\eta $ be a tensor, so there arises the issue as to whether the constraint can make sense  globally if $\eta$ is a generalised tensor rather than a tensor. 
A natural generalisation is to consider versions of
DFT in which  $\eta$ is replaced by a general (non-constant) metric of signature $(D,D)$, and this has been explored in \cite{Cederwall:2014kxa}.  Here we will restrict ourselves to  the case of constant $\eta$.

If the doubled space involves a doubled torus or a bundle  with doubled torus fibres,
then contact can be made with string theory on a torus or T-fold, and the significance of the doubled geometry is that explained in  \cite{Hull:2004in}. However, the background independent formulation of  \cite{Hohm:2010jy,Hohm:2010pp}  suggests DFT might be written on more general doubled spaces $ M$, 
constructed from local patches of the kind considered above.
 This leads to interesting questions as to the geometry and significance of $ M$.  
  For   non-toroidal doubled spaces $ M$, there is the question of the meaning (if any) of the extra coordinates $\ti x$.  For a general space-time $N$, there need not be any winding modes, or the number of winding modes (given by the number of topologically distinct non-contractible loops) might be different from the number of  momenta, so that 
  for general spaces
  there will not be expected to be any  T-duality and the $\ti x$  cannot be associated with winding modes.

To better understand the  geometry of DFT, a number of attempts have been made to explore the
relationship between the gauge symmetries of DFT and the diffeomorphisms of the doubled space. 
Despite a number of formal  similarities (e.g. one acts through Lie derivatives, the other through generalised Lie derivatives) the gauge group and the diffeomorphism group are not isomorphic.
The    DFT gauge transformations of 
\cite{Hohm:2010jy,Hohm:2010pp} 
act on fields  at a point $X\in   M$, taking fields at $X$ to transformed fields at $X$,  $A(X)\to A'(X)$.
Just as diffeomorphisms can be written in either an active or a passive form, it is natural to ask whether the DFT transformations could be written in a form in which the coordinates $X$ transform.
This could be helpful in understanding finite gauge transformations, analysing the patching together of different regions of the doubled space, and in addressing the question  of whether duality transformations can be understood as arising from   gauge transformations.

Expressions for gauge transformations with finite parameters in which fields transform at a point $X$, 
$A(X)\to A'(X)$, are obtained by expoentiating the infinitesimal transformations of \cite{Hohm:2010jy,Hohm:2010pp}.
An alternative  form for finite gauge transformations that acts on the coordinates has been proposed recently  in \cite{Hohm:2012gk}, and then   related forms of this were developed in  
 \cite{Park:2013mpa} and \cite{Berman:2014jba}.  
The proposal of \cite{Hohm:2012gk} 
gives transformations with a non-associative composition rule
and 
 it was suggested in \cite{Hohm:2013bwa}
that this leads to a kind of non-associative geometry. 
This appears to be in tension with the formulation of double geometry of \cite{Hull:2004in} in which the doubled space is a conventional manifold, and with the fact that the DFT is locally equivalent to 
a
(super)gravity theory that has 
gauge transformations that compose associatively. The use of these transformations as transition functions was considered in \cite{Hohm:2013bwa,Papadopoulos:2014mxa}.

In  \cite{Park:2013mpa} it was suggested that physical points should correspond
not to points in the doubled space $M$ but 
 to orbits in $  M$ under transformations referred to as \lq coordinate gauge symmetries'. Forms for finite gauge transformations with $X^M$ transforming
 were proposed and shown to give the correct results modulo certain DFT gauge symmetries.
 In  the formulation  of \cite{Berman:2014jba}, the doubled space is a conventional manifold, as in \cite{Hull:2004in}, and
 a key role was played by certain local
$O(D,D)$ transformations. In \cite{Berman:2014jba},  forms for finite gauge transformations with $X^M$ transforming
 were proposed and shown to give the correct results modulo the local
$O(D,D)$ transformations. 
However, as   \cite{Park:2013mpa} and \cite{Berman:2014jba} give forms of finite gauge transformations (with $X$ transforming) only up to certain DFT gauge symmetries or local
$O(D,D)$ transformations, they lose track of an important part of the finite gauge transformations. As will be seen, they  only encode the  diffeomorphisms $x^m\to x^m + \xi ^m  (x)+\dots$ of the subspace with coordinates $x$, and lose almost all information about the anti-symmetric tensor gauge transformations.

These three proposals attempt to represent  finite DFT gauge transformations in terms of transformations that act as coordinate transformations on the coordinates $X\to X'(X)$. This cannot be an isomorphism as the DFT gauge group and the diffeomorphisms of the doubled space are different groups.
In  \cite{Hohm:2012gk}, it was proposed  to resolve this by introducing a new star product composition of 
coordinate transformations of $  M$, which turns out to be non-associative. 
In  \cite{Berman:2014jba}, it was argued instead  that the finite gauge transformations 
used there provide a homomorphism
 up to  local
$O(D,D)$ transformations. This leads to three elements $g_1,g_2, g_3$ of the group of  finite DFT gauge transformations with $g_1g_2g_3=1$ being represented not by the identity transformation but
by a \lq cocycle'  that is a  local
$O(D,D)$ transformation. This was then argued to reveal an underlying gerbe structure of the doubled manifold \cite{Berman:2014jba} when such gauge transformations were used as transition functions.

The aim of this paper is to a present a new  explicit and simple  form for finite gauge transformations in DFT in which the coordinates transform and which avoid some of the issues arising with other approaches that were outlined above. They are consistent with $  M$ being a conventional manifold, they are associative and they agree with the forms obtained by exponentiating infinitesimal transformations exactly, not just modulo 
 coordinate gauge symmetries or local
$O(D,D)$ transformations. 
They elucidate the relationship with generalised geometry \cite{Hitchin:2004ut, Gualtieri:2003dx, Hitchin:2005in,Hitchin:2010qz}
and reveal an explicit gerbe structure.
Here the main focus will be on  DFT in a local patch  ${\cal U}$ with constant metric $\eta$, and   the transition functions and global structure will be addressed  in 
 a separate paper.

\section{Double Field Theory }\setcounter{equation}{0}

Double field theory  is formulated in a doubled space-time $  M$ with coordinates $X^M$
where $M,N=1,\dots 2D$ and a constant $O(D,D)$ invariant \lq metric'
 $\eta_{MN}$, which is used to raise and lower indices. The indices $M,N,\dots$ transform covariantly under $O(D,D)$, so that e.g. $V^M$ is an $O(D,D)$ vector, while $V^MW_M$ is $O(D,D)$ invariant.
  If the fields and  parameters of gauge transformations are required to satisfy 
 the   `strong constraint'  (so that $\partial^M\partial_M A = 0$  
 and $\partial^M A\, \partial_M B=0$
for any fields or parameters $A$ and $B$),
then the theory is locally equivalent to the standard 
theory of metric, $b$-field and dilaton.
The strong constraint 
 implies \cite{Hohm:2010jy}  that locally  all fields depend on only $D$ of the coordinates, 
 and these parameterise a subspace of a coordinate patch that is null with respect to  
 $\eta$.
 
The theory can be formulated  \cite{Hohm:2010pp} in terms of a generalised metric ${\cal H}_{MN}$ which encodes the metric $g_{mn}$ and  2-form gauge field $b_{mn}$, together with a scalar  density $d$.
The theory has an infinitesimal symmetry
 \be\label{gaugevar}
 \begin{split}
  \delta_{\xi}{\cal H}_{MN} \ &= \ 
  \xi^{P}\partial_{P}{\cal H}_{MN}+\big(\partial_{M}\xi^{P}-\partial^{P}\xi_{M}\big)
  {\cal H}_{PN}+\big(\partial_{N}\xi^{P}-\partial^{P}\xi_{N}\big)
  {\cal H}_{MP}\;, \\
  \delta_{\xi}d \ &= \ \xi^{M}\partial_M d-\frac{1}{2}\partial_{M}\xi^{M}\;
 \end{split}
 \ee  
 with an    $O(D,D)$ vector parameter $\xi^{M}(X)$.
 The strong constraint on all fields and parameters is used in proving gauge invariance of the action of \cite{Hohm:2010pp}.

\subsection{Generalised Lie Derivatives}

The gauge transformation of the  generalised metric can be written in terms of a generalised
Lie derivative \cite{Hohm:2010pp}:   
$$\delta_{\xi}{\cal H}_{MN} =  \widehat{\cal L}_{\xi}{\cal H}_{MN}$$
 A {\it generalised tensor
} $T^{M\dots N}{}_{P\dots Q}$ is defined as  transforming under the DFT gauge transformations via the generalised
Lie derivative, $\delta _\xi T^{M\dots N}{}_{P\dots Q}= \widehat{\cal L}_{\xi}T^{M\dots N}{}_{P\dots Q}$,
so that the
 generalised metric is a generalised tensor.
The generalised Lie derivative of  a generalised tensor $A_M$ with one lower index   is
 \be\label{genLievec}
  \widehat{{\cal L}}_{\xi} A_{M} \ = \ \xi^{P}\partial_{P}A_{M}{}
  +(\partial_{M}\xi^{P}
  \, -\partial^{P}\xi_{M})\,A_{P}\;,
 \ee
 while for a generalised tensor $A^M$ with an upper index   it   is
 \be\label{genLievecup}
  \widehat{{\cal L}}_{\xi} A^{M} \ = \ \xi^{P}\partial_{P}A^{M}{}
  +(\partial^{M}\xi_{P}
  \, -\partial_{P}\xi^{M})\,A^{P}\;.
 \ee
 This then extends to arbitrary tensors using the Lebnitz rule and linearity \cite{Hohm:2010pp}.
The  
generalised Lie  derivatives of the $O(D,D)$ metric $\eta_{MN}$
and the Kronecker tensor $\delta_M{}^N$ vanish:
\be
 \widehat{\cal L}_{\xi}\eta_{MN}=0,\qquad
  \widehat{\cal L}_{\xi}\delta _M{}^N=0\;.
  \ee

The generalised Lie derivative
$ \widehat{{\cal L}}_{\xi} $ of {\em any} generalised tensor
vanishes when $\xi^M = \partial^M \chi$, so that for
any generalised tensor $T$ satisfying the strong constraint, we have  
\be
\label{varH}
\widehat{\cal L}_{\xi+ \eta^{-1} \partial\chi}T= \widehat{\cal L}_{\xi}T\;.
\ee
We shall refer to transformations with parameter of the form $\xi^M_{red} = \partial^M \chi$ 
with $ \widehat{{\cal L}}_{\xi_{red}} =0$
as redundant transformations.
 
 The commutator  of two generalised Lie  derivatives is
  \be
  \label{bracLie}
  \big[\widehat{\cal L}_{\xi_1} , \widehat{\cal L}_{\xi_2} \big] \ = \  
    \widehat{\cal L}_{[\xi_1, \xi_2]_{{}_C}}\;, 
 \ee
with the C-bracket \cite{Siegel:1993th,Hull:2009zb} 
 \be\label{Cbracket}
   \bigl[ \xi_1,\xi_2\bigr]_{{}_C}^M  
   \ \equiv
   \ \xi_{1}^{N}\partial_{N}\xi_{2}^M -\frac{1}{2}\, 
    \xi_{1N}\partial^{M}\xi_{2}^N
   -(1\leftrightarrow 2)\; .
 \ee
 This is an $O(D,D)$  covariant form of the Courant bracket \cite{Hull:2009zb}.
 Then the gauge algebra is
  \be\label{Cbracket}
  \big[\delta_{\xi_1},\delta_{\xi_2}\big] \ =
   \delta_{\,[\xi_1,\xi_2]_c} 
 \ee
The C-bracket does not satisfy the Jacobi identity \cite{Hull:2009zb}:
 \be
\big[\xi_1,\big[\xi_2,\xi_3\big]_{{}_C}\big]_{{}_C}+{\rm cyclic} \ = \ 
J(\xi_1,\xi_2,\xi_3) 
\ee
where the Jacobiator is $J^M=\partial^M {\cal N} $,
and $ {\cal N} $ is the Nijenhuis tensor defined by 
\be\label{Nijenhuis}
  {\cal N} (\xi_1,\xi_2,\xi_3) \ = \  \frac{1}{6}\Big(\big\langle \big[\xi_1,\xi_2\big]_C,\xi_3\big\rangle
  +{\rm cyclic}  \Big)\;. 
\ee  
As  $J^M=\partial^M  {\cal N} $ parameterises a redundant gauge transformation,  it follows that
$$
  \widehat{{\cal L}}_J A=0
  $$
  for any tensor $A$ satisfying the strong constraint.
  Then the generalised Lie  derivatives satisfy the Jacobi identity 
    \be
 \big[  \big[\widehat{\cal L}_{\xi_1} , \widehat{\cal L}_{\xi_2} \big] , \widehat{\cal L}_{\xi_3} 
 \big]  +{\rm cyclic}
 \ = \  0
  \ee
  which is essential for the gauge transformations given in terms of the generalised Lie  derivative to be a symmetry \cite{Hull:2009zb}.
  
 The C-bracket can be written
 as
  \be\label{Cbracket}
   \bigl[ \xi_1,\xi_2\bigr]_{{}_C}^M  
   \ = \
  \bigl[ \xi_1,\xi_2\bigr]{{}}^M  + \l^M_{12}, \qquad 
  \lambda  ^M_{12}\equiv
  -\frac{1}{2}\, 
    \xi_{1N}\partial^{M}\xi_{2}^N
   -(1\leftrightarrow 2)\;, 
 \ee
 where $  \bigl[ \xi_1,\xi_2\bigr]$  is the ordinary Lie bracket on the doubled space.  
This was used in \cite{Berman:2014jba} to write the algebra (\ref{bracLie}) as  
  \be
  \label{bracLieD}
  \big[\widehat{\cal L}_{\xi_1} , \widehat{\cal L}_{\xi_2} \big] \ = \  
    \widehat{\cal L}_{[\xi_1, \xi_2]}  + \Delta _{12}
 \ee
 where 
 \be
 \label{ahdf}
  \Delta _{12}=\widehat{\cal L}_{\lambda _{12}}
  \ee
  In \cite{Berman:2014jba}, it was emphasised that the gauge transformation $\widehat{\cal L}_{\lambda _{12}}$
   involves no translation term when acting on tensors $T$ satisfying the strong constraint as
   $\lambda ^M \partial _M T=0$, and can be viewed as a local $O(D,D)$ transformation.
   Such 
\lq $\D$-transformations' played a key role in the construction of \cite{Berman:2014jba}, and will be discussed further in later sections.
 
\subsection{Solving the Strong Constraint}\label{StrongResolve}

It was shown in  \cite{Hohm:2010jy}  
 that the strong constraint implies that,
at least locally, all fields are restricted to a $D$-dimensional null subspace.
Consider then a patch $\cal U$
 of $  M$, which is diffeomorphic to a patch of $\R^{2D}$ with coordinates $X^M$ and constant metric $\eta$. Then the strong constraint implies that the DFT fields only depend on the coordinates of a totally null subspace  $U\subset {\cal U}$  \cite{Hohm:2010jy,Hohm:2013bwa}. Let the coordinates of $U$ be $x^m$ and the remaining coordinates be $\ti x_m$, so that 
 \be
X^M = \begin{pmatrix} x^m \\   \tilde x_m \end{pmatrix} \, , \qquad
\partial_M =   \begin{pmatrix} \partial_m\\  \tilde \partial^m  \end{pmatrix}
\ee
and the $O(D,D)$ invariant metric is
\be
\label{etais}
\eta_{MN} =  \begin{pmatrix}
0&1 \\1&0 \end{pmatrix}\,.
\ee
In this coordinate basis, a generalised vector then decomposes as
\be
\xi ^M = \begin{pmatrix}   \xi ^m \\ \tilde \xi _m \end{pmatrix} \, ,
\ee
while the generalised metric takes the form
\be\label{His}
  {\cal H}_{MN} \ = \  \begin{pmatrix}    
  g_{mn}-b_{mk}g^{kl}b_{ln}  & 
 b_{mk}g^{kn}   \\[0.5ex]
 -g^{mk}b_{kn}   &
  g^{mn}
   \end{pmatrix}   
   \;
 \ee
in terms of the metric $g_{mn}$ and antisymmetric tensor gauge field $b_{mn}$.

The strong constraint is solved in the patch $\cal U$ by having
 all fields and parameters independent of $\tilde x_m$ so that
\be
 \tilde \partial^m  =0
 \ee
 on all fields.
 Then the fields and parameters    depend only on the coordinates $x^m$, parameterising the $D$-dimensional patch $U\subset {\cal U}$, so can be regarded as fields on the  totally null subspace $U$.
It was shown in 
 \cite{Hohm:2010pp} that the transformation (\ref{varH}) then becomes
\be
 \label{nonlingauge}
 \begin{split}
  \delta_{\xi}g_{ij} &= {\cal L}_{\xi}g_{ij}
  \\[0.8ex]
  \delta_{\xi}b_{ij} &= {\cal L}_{\xi}b_{ij}+{\cal L}_{\tilde{\xi}}b_{ij}
  +\partial_{i}\tilde{\xi}_{j}-\partial_{j}\tilde{\xi}_{i}
  \end{split}
 \ee
so that the $\xi^m(x)$ are the parameters of diffeomorphisms acting through the ordinary Lie derivative ${\cal L}_{\xi}$ and $\tilde \xi_m(x)$ are the parameters of antisymmetric tensor gauge transformations.
 
 Note that the coordinates $x^m$  on which the fields depend need  not be the physical space-time coordinates. A choice of {\it polarisation} splits the $2D$ coordinates $X^M$ into $D$ space-time coordinates and $D $ dual winding coordinates 
 \cite{Hull:2004in}.  This 
 choice of splitting changes under T-duality   and need not 
correspond to the splitting into the coordinates $x$ on which the fields depend, and  the remaining coordinates $\tilde x$.
 However, it was shown in  \cite{Hohm:2010jy}
  that one can always choose a polarisation or duality frame in which, for a given patch,  the coordinates $x^m$ are the coordinates for a patch of space-time  and the $\tilde x_m$ are the corresponding  winding coordinates. 
   
\sectiono{Reducibility and the Symmetry Group }\label{Jacalg} \setcounter{equation}{0}

In this section, we analyse the algebraic structure underlying the symmetries of DFT further, following the approach of  \cite{Hull:2009zb}.
The parameters $\xi ^M(X) $ can be written formally as $\xi ^A $ where $A$ is 
 a composite index representing the discrete index $M$ and the continuous variables $X$,  with summation over $A$ representing summation over  $M$ and integration over  the coordinates  $X$.
 The C-bracket defines constants
 $f_{AB}{}^C$ by
 \be
 \bigl( [\xi_1, \xi_2]_C \bigr)^A= - 2f_{BC}{}^A \xi_1^B\xi_2^C
  \ee
 These can then be used as structure constants for a 
 closed algebra ${\cal K}$ with formal generators $T_A$ 
\be
\label{alg}
[T_A, T_B]=f_{AB}{}^CT_C\,.
\ee
This is not a Lie algebra, as there is a non-trivial Jacobiator
\be
[[T_A,T_B],T_C] + {\rm cyclic~ permutations}=
g_{ABC}{}^D T_D  
\ee
given by constants $g_{ABC}{}^D =-3f_{[AB}{}^Ef_{C]E}{}^D $.

The redundant transformations  with parameters $\xi^M =\partial ^M\chi$   form an invariant subalgebra.
That is, we can choose a basis of   generators
$T_A=\{ t_a, Z_\a \}$ where $Z_\a$ generate the redundant transformations, and the $t_a$ are
a basis for the remaining generators.
The $Z_\a$ generate an invariant subalgebra $\cal Z$ 
 so that 
 the algebra is of the form
 \be
 [T_A, Z_\a]= f_{A\a}{}^\b Z_\b,\qquad [t_a, t_b] = f_{ab}{}^ ct_c
 +  f_{ab}{}^\g Z_\g
  \ee
  Moreover, the Jacobiator is in $\cal Z$:
 \be
[[T_A,T_B],T_C] + {\rm cyclic~ permutations}=
g_{ABC}{}^\alpha Z_\alpha  \,.    
\ee
The quotient
${\cal K}/{\cal Z}$ defines a Lie algebra $\bf h$ with structure constants $ f_{ab}{}^ c$,
  as the   $ f_{ab}{}^ c$ satisfy the Jacobi identities.
It will be useful to make a corresponding split of the parameters, so that
\be
\xi^AT_A= \r ^a t_a+  \zeta ^\a Z_\a
\ee

Suppose one were to attempt to define a linear representation of $\cal K$  in which $T_A$ is represented by a linear transformation $L(T_A)$,
with the commutators of linear transformations satisfying
\be
\label{Lalg}
[L(T_A), L(T_B)]=f_{AB}{}^CL(T_C)\,.
\ee
This will fail in general as commutators of linear transformations   satisfy the Jacobi identity while the structure constants $f_{AB}{}^C$ do not.
However, such a representation
 can be consistently defined if the generators $Z_\a$ are represented trivially, $L(Z_\a)=0$, so that it is a representation of the quotient ${\bf h}={\cal K}/{\cal Z}$.
Then we require $L$ to satisfy
\be
L(\xi^AT_A)=\r^a L(t_a)
\ee
and
\be
[L(t_a), L(t_b)] = f_{ab}{}^ cL(t_c)
\ee
so that the $L$ provide a representation of the lie algebra ${\bf h}$.
For finite parameters,
exponentiating then gives finite  transformations
\be 
h(\xi)\equiv \exp L(\xi^AT_A) = \exp 
 \r^a L(t_a)
 \ee
 which are elements of a Lie group H that has Lie algebra  ${\bf h}$.
 The generalised Lie derivatives provide just such a representation acting on generalised tensors, with $\xi^A L(T_A)= {\widehat {\cal L}} _\xi$.
The finite transformations given by exponentiation
 gives   the symmetry group $H$ of double field theory (with the strong constraint).
 
The group of gauge transformations   then has the composition
\be
h(\xi_1) h(\xi_2) = h(\xi_{12})
\ee
where for infinitesimal parameters
 \be \label{xicomm}
  \xi_{12}=\xi_1+\xi_2 - \frac 1 2
[ \xi_1, \xi_2 ]_C +\dots
\ee
As for each $\xi =(\r,\z)$, only the $\r$ part acts,
we can write
$h(\xi) =h(\r)$ and find
\be
h(\r_1) h(\r_2) = h(\r_{12})
\ee
with the Lie group multiplication giving $\r _{12}$ via the Baker-Campbell-Hausdorff formula:
\be
\r_{12}^a= \r_1^a+\r _2^a - \frac 1 2  f_{bc}{}^ a \r^b _1\r^c_2 +\dots
 \ee

\section{Review of Proposals for Finite Transformations}\setcounter{equation}{0}

For diffeomorphisms of a manifold with coordinates $x^m$, a tensor field $T$ transforms via the standard Lie derivative with respect to a vector field $v^m(x)$:
\be
\delta T = {\cal L}_v T
\ee
This generates a finite transformation via exponentiation:
\be
 T'(x)= e^{ {\cal L}_v} T (x)
 \ee
 A   useful form of the transformation can be given by  rewriting in terms of a change of coordinates
\be
x\to x'(x), \qquad x'= e^{-v^m\partial _m} x
\ee
For example, for a covector $T_m$,
the transformation becomes
\be
T'_m(x') = T_n(x) \frac{\partial x^n } {\partial x'{}^m}
\ee

Similarly, for the gauge transformations of DFT,
a finite transformation is   given by   exponentiating the generalised Lie derivative. For example, for a generalised 
tensor $T_M$, 
 \be\label{expl}
  T_{M}^{\prime}(X) \ = \ e^{\widehat{\cal L}_{\xi}} \,T_{M}(X)\;, 
 \ee  
where 
all fields and parameters 
depend on $X$ and satisfy the strong constraint.  
This exponentiation has been studied in \cite{Hohm:2012gk,Park:2013mpa,Berman:2014jba} where explicit expressions for the finite transformations have been rewritten in various forms using the strong constraint.

In \cite{Hohm:2012gk}, the question was raised as to whether there was a useful way of rewriting this in terms of a transformation of the doubled coordinates $X^M$, 
$X\rightarrow X^{\prime} = f(X)$.
In \cite{Hohm:2012gk}, 
 the following
transformation for an 
$O(D,D)$ vector $T_{M}$ was  proposed: 
 \be\label{finiteF}
   T_{M}^{\prime}(X^{\prime}) \ = \ {\cal F}_M{}^N   T_N(X)\,,
   \ee
where the matrix ${\cal F}$ is defined by
   \be\label{Fdef} \phantom{\Biggl(}
 {\cal F}_M{}^N\ \equiv \   {1\over 2} 
  \Bigl(   \frac{\partial X^{P}}{\partial X^{\prime M}}\,
  \frac{\partial X^{\prime}_P}{\partial X_{N}}
  + \frac{\partial X'_{ M}}{\partial X_P}\,
  \frac{\partial X^N}{\partial X^{\prime P}}\Bigr)
  \,. 
 \ee
Here 
the indices on coordinates are raised and lowered with  $\eta$.
A tensor with an arbitrary number of   indices transforms `tensorially', 
with each index rotated by the matrix ${\cal F}$.   It was shown in \cite{Hohm:2012gk}
 that
${\cal F}$ is in fact an $O(D,D)$ matrix, so that $\eta_{MN}$ is invariant.
This is different from a coordinate transformation on a cotangent vector field  of the doubled space, for which there would be a similar transformation with 
 ${\cal F}_M{}^N$ replaced by ${\partial X^N\over \partial X'^M}$; the metric $\eta_{MN}$ would not in general be invariant under such coordinate transformations.

However, this proposal doesn't quite work if $X'$ is given by the expected transformation
\be
\label{xtranss}
X'^M \ = \  e^{- \xi^K \partial_K}  X^M 
\ee
In particular, it doesn't reproduce the transformation (\ref{expl}), and for transformations $X\to X'\to X''$, it doesn't have the desired property 
\be
\label{compose}
{\cal F} (X'' , X') \, {\cal F} (X' , X) = {\cal F} (X'', X)\, .
\ee
In \cite{Hohm:2012gk}, it was proposed that instead $X'(X)$ should be given by
\be
\label{xprimeis}
X'^M \ = \  e^{-\Theta^K(\xi)\partial_K}  X^M \,, ~~~~
\Theta^K(\xi) \, \equiv  \ \xi^K   + {\cal O} (\xi^3)\,,
\ee
and $\Theta^K(\xi)$ was found
to ${\cal O}(\xi^4)$.  The results of \cite{Berman:2014jba} effectively determine $\Theta^K(\xi)$ to all orders.
With this form of $X'$, the transformations (\ref{finiteF}) were shown to give the same result for $T'$ as   (\ref{expl}), and the composition law (\ref{compose}) was shown to hold \cite{Hohm:2012gk}.

It is important for the approach of \cite{Hohm:2012gk}
 that  $\Theta $ has a form given by 
\be
\Theta^M \ = \ \xi^M +  \sum_{i}    
\rho_{i}\,
\, \partial^M \chi_{i}\,,
\ee
with $\rho_i$ and $\chi_i$ functions of $\xi$ and $X$, so that, when acting on fields satisfying the strong constraint,
\be
\Theta^P \partial_P
=\xi^P \partial_P
\ee
and
\be \widehat{{\cal L}}_{\Theta(\xi)} =  \widehat{{\cal L}}_{\xi} 
\ee

It will be convenient to denote the transformation with finite parameter $\xi$ by $k(\xi)$, so that (\ref{finiteF}) can be written
$T'= k(\xi)T$.
Under composition, these would combine in the natural way
to give
\be
k(\xi_{12}) T = k(\xi_1)\Bigl( k(\xi_2) T\Bigr)
\ee
which would imply \cite{Hohm:2012gk}
\be \label{xicommc}
  \xi_{12}=\xi_1+\xi_2 - \frac 1 2
[ \xi_1, \xi_2 ] +\dots
\ee
with the ordinary Lie bracket. This is different from the composition law for DFT gauge transformations (\ref{xicomm}) which is of   similar form, but with the C-bracket instead of the Lie bracket.
In \cite{Hohm:2012gk,Hohm:2013bwa}, it was proposed that the multiplication of  these transformations be modified to a \lq star product' $k_1\star k_2$ with
\be
\label{starstar}
k(\xi_{12})  = k(\xi_1)\star k(\xi_2)  
\ee
with $\xi_{12}$ now given to lowest order by
 \be  \xi_{12}=\xi_1+\xi_2 - \frac 1 2
[ \xi_1, \xi_2 ]_C +\dots
\ee
 so that it is 
 determined by the C-bracket. It was conjectured that this could be done to all orders, so that the gauge algebra of transformations would be   consistent with that of DFT.
 
 However, this star product is not associative
\be
(k_1\star k_2) \star k_3 \ne
 k_1\star (k_2 \star k_3 )
 \ee
 The violation of associativity is determined  to lowest order in infinitesimal parameters  by the Jacobiator. Indeed, an explicit calculation in \cite{Hohm:2013bwa}
 implies
 \be
\Bigl[(k_1\star k_2) \star k_3  \Bigr] 
\star
\Bigl[
 k_1\star (k_2 \star k_3 ) \Bigr] ^{-1} = 
 k (\xi_J), \qquad \xi _J\equiv - \frac 1 6 J(\xi_1,\xi_2,\xi_3) +O(\xi^4)
 \ee
 where $k_i=k(\xi_i)$.
 In \cite{Hohm:2013bwa}, it was suggested that this non-associativity of the product of transformations could have an interpretation in terms of a non-associative geometry for doubled space-time.
 
The transformations $k(\xi)$ have the property that they are non-trivial for the parameters $\xi^M=\partial^M\chi$
of redundant gauge transformations, $k(\partial^M\chi)\ne 1$. The Jacobiator is of the form 
$J=\partial^M 
 {\cal N} 
$ and defines a non-trivial transformation $k(J)$, leading to the failure of 
associativity.  

In \cite{Berman:2014jba}, a variant on this construction was proposed.
The transformation (\ref{finiteF}) with (\ref{Fdef}) was again used but now with the standard transformation for 
$X$ under a diffeomorphism generated by $\xi$, given by
\be
\label{xtranssa}
X'^M \ = \  e^{- \xi^K \partial_K}  X^M \, .
\ee
This no longer reproduced the transformation (\ref{expl}) or satisfied the composition law (\ref{compose}), 
as the transformation of $X$ is different from that of \cite{Berman:2014jba}. However, in \cite{Berman:2014jba}
 they showed that this transformation
agrees with (\ref{expl}) up to $\Delta$-transformations, the local $O(D,D)$ transformations arising in (\ref{bracLieD}),
and the  composition law (\ref{compose}) is satisfied up to $\Delta$-transformations. This formulation requires no non-associativity, but was argued to involve a gerbe-like structure on doubled space-time.

In \cite{Park:2013mpa}, it was pointed out that for any field $T(X)$ satisfying the strong constraint,
$T(X+\r) = T(X)$
for any
$\r$ of the form
$\r ^M = \phi \partial ^M \chi$ for some $\phi(X), \chi (X)$.
This was referred to as a \lq coordinate gauge symmetry' and it was proposed that physical points should correspond to gauge orbits under the transformations
\be
\label{coordgauge}
X^M\to X^M+  \phi \partial ^M \chi
\ee
The coordinate gauge transformations were then associated with 
DFT gauge transformations with parameter $\rho ^M$.
This gives a similar picture to \cite{Berman:2014jba}:
the transformation (\ref{finiteF}) with (\ref{Fdef})  with (\ref{xtranssa}) agrees with (\ref{expl}) up to such
 DFT gauge transformations associated with coordinate gauge transformations.

\section{Discussion of Proposals for Finite Transformations} \setcounter{equation}{0}

\subsection{The Non-Associative Proposal}

The double field theory gauge transformations  $h(\xi)$ with finite parameters $\xi$ represent elements of a Lie group $H$.
In  \cite{Hohm:2012gk}, the effect of any given gauge transformation on a generalised tensor is reproduced by a transformation of the form (\ref{finiteF})  consisting of (i) a transformation $X\to X'(X)$ and (ii) a local $O(D,D)$ transformation on each tensor index.
Here we wish to focus on the transformation $X\to X'(X)$, which for \cite{Hohm:2012gk} is given by (\ref{xprimeis}).
Then the proposal of \cite{Hohm:2012gk}  gives a map  $\phi$ from 
 the set  ${\rm {Diff }}(  M)$ of diffeomorphisms of the doubled space, to $H$:
\be
\phi: {\rm {Diff }}(  M) \to H
\ee
with
\be 
\phi: d (\xi)=e^{-\Theta^K(\xi)\partial_K}  \to 
h(\xi) = e^{{\widehat {\cal L}} _\xi}
\ee
Note that this  map $\phi$ is not invertible, as there is a non-trivial kernel consisting of diffeomorphisms $d(\xi)$ with parameter  of the form $\xi ^M=\partial ^M\chi$.

The diffeomorphisms ${\rm {Diff }}(  M)$ have a standard group structure given by composition $d_1\cdot d_2$, so that $(d_1\cdot d_2) f = d_1(d_2 f)$ for any function $f(X)$ and $d_1,d_2 \in {\rm {Diff }}(  M) $.
The map $\phi$ is not a homomorphism:
\be 
\phi (d_1 \cdot  d_2)\ne  \phi (d_1 ) \phi ( d_2)
\ee
The approach of \cite{Hohm:2012gk,Hohm:2013bwa} attempts to define a star product for elements of ${\rm {Diff }}(  M)$ that makes this a homomorphism:
\be 
\phi (d_1 \star  d_2)=  \phi (d_1 ) \phi ( d_2)
\ee
The idea is that this should give a realisation of the DFT gauge transformations as diffeomorphisms of the doubled space. Note that as $\phi$ has a non-trivial kernel, this requirement does not determine the star product completely. However, it determines it up to redundant gauge transformations.  The ambiguity can be largely fixed by requiring $O(D,D)$ covariance, which gives the star product of \cite{Hohm:2012gk}, but this choice has the drawback of giving a non-associative multiplication.
However, this construction raises a number of issues as ${\rm {Diff }}(  M)$ and $H$ are different groups, and not homomorphic.

To illustrate the issues, consider two different Lie groups $G,G'$ of the same finite dimension $d_G$, with generators $T_a,T'_a$ in the corresponding Lie algebras, $a=1,\dots d_G$.
For example, we might take $G=GL(3,\R)$ and $G'=SU(2)\times SU(2)\times SU(2)$.
Then $G$ will contain elements of the form $g=e^{\xi^aT_a}$ and  $G'$ will contain elements of the form $g'=e^{\s ^aT'_a}$. One can then define a  map $\phi :G\to G'$ between exponential group elements by
\be
\phi: g=\exp( {\xi^aT_a})
\to \phi(g)=\exp( {\xi^aT'_a})
\ee
or more generally by
\be
\phi: g=\exp( {\xi^aT_a})
\to \phi(g)=\exp( {f(\xi)^aT'_a})
\ee
with
$f(\xi)^a$ an invertible (and possibly non-linear) map  $\R^{d_G}\to \R^{d_G}$.
This is   not a homomorphism, but is an invertible map on the exponential group elements.
One could attempt to define a new star product on $G$ that made it a homomorphism:
\be
g_1 \star g_2 = \phi ^{-1} \left( \phi (g_1) \cdot \phi (g_2) \right)
\ee
where $g'_1\cdot g'_2$ is the $G'$ group multiplication.
This would mean trying to impose a $G'$ multiplication rule on elements of $G$.
There are of course a number of problems with such an attempt.
For Lie groups, the algebraic structure of the Lie algebra determines much of the geometry, and it is inconsistent to impose the wrong multiplication rule on a given geometry.
Not all elements of the groups $G,G'$ will be of exponential form in general, and there will be problems with extending the map $\phi$ smoothly to non-exponential group elements. If one has a set with the multiplication rules of $G'$, then one is really dealing with the Lie group $G'$, not $G$.

Consider now a similar set-up, but with $G$ of greater dimension than $G'$, $d_G> d_{G'}$, and generators $T_a$ of $G$ and $T'_\a$ of $G'$, $\a=1,\dots d_{G'}$.
For example, we might take $G=GL(3,\R)$ and $G'=SU(2)\times SU(2)$.
We can consider a map from exponential elements of $G$ to exponential elements of $G'$
with
\be
\phi: g=\exp( {\xi^aT_a})
\to \phi(g)=\exp( {f(\xi)^\a T'_\a } )
\ee
where
$f(\xi)^a$ is a (possibly non-linear) map  $f:\R^{d_G}\to \R^{d_{G'}}$. 
This would not be a homomorphism in general, but one again 
could attempt to 
define a new star product on $G$ that made it a homomorphism by requiring:
\be
\phi 
 \left( g_1 \star g_2 \right)=\phi (g_1) \cdot \phi (g_2) 
\ee
As $\phi $ is no  longer invertible, this does not completely determine the star product.
However, it will be imposing a product on $G$  that is partially determined by the product in $G'$, and similar objections to those above would again hold.

The construction of \cite{Hohm:2012gk,Hohm:2013bwa} is    similar to these examples, trying to impose the multiplication of $H$ on the group  ${\rm {Diff }}(  M)$. 
The symmetries of DFT are not diffeomorphisms of the doubled space $ M$ and 
have a different group structure from the diffeomorphisms.  Any attempt to realise DFT gauge transformations in terms of transformations $X\to X'(X)$ is likely to  be problematic.

\subsection{The Proposal with Local $O(D,D)$}

Consider now the proposal of \cite{Berman:2014jba}. We again focus on the transformation of $X$.  In \cite{Berman:2014jba}, 
the coordinate transformation $X\to X'(X)= d(\xi)X$ where 
\be
d (\xi)=e^{-\xi^K\partial_K}  
\ee
of the doubled space 
is associated with the DFT gauge transformation
\be 
h(\xi) = e^{{\widehat {\cal L}} _\xi}
\ee
This map $d(\xi)\to h(\xi)$ cannot be a homomorphism from ${\rm {Diff }}(  M)$ to $H$. This and related issues are dealt with by the authors of  \cite{Berman:2014jba} by working modulo the $\Delta$ transformations arising in the algebra (\ref{bracLieD}), which  they  refer to as non-translating local $O(D,D)$ transformations.
They show that the transformations resulting from $d(\xi)$ and $h(\xi)$ agree modulo such local $O(D,D)$ transformations, and that the composition rules also agree
up to 
such
transformations.
Thus the map from ${\rm {Diff }}(  M)$ to $H$ might be thought of as a \lq homomorphism
 up to local $O(D,D)$ transformations'.
 In \cite{Berman:2014jba}, 
 it was proposed also  that the local $O(D,D)$ transformations were the key to resolving a number of issues in DFT.
 
To understand this further,  we will now investigate
these local $O(D,D)$ transformations
 in a patch in which the fields and parameters depend on $x^m$ but not $\ti x_m$, as in section \ref{StrongResolve}.
The $\lambda^M= (\lambda ^m,\tilde \lambda _m)$ defined in 
(\ref{Cbracket})
then takes the form 
\be
  \lambda ^m_{12}=0
  ,\qquad 
  \ti \lambda_m{}_{12}=  -\frac{1}{2}\,
    \xi_{1N}\partial_m\xi_{2}^N
   -(1\leftrightarrow 2)\;, 
 \ee
 The transformation $ \Delta _{12}=\widehat{\cal L}_{\lambda_{12}}$
 is then just the anti-symmetric tensor gauge transformation 
 with parameter $ \ti \lambda_m{}_{12}$, giving $\d b_{mn}=\partial _{[m}\ti \lambda_{n]} $ and can be written in terms of the action on the generalised metric of 
 the infinitesimal $O(D,D)$ matrix
 \be
\label{riss}
\Delta =  \begin{pmatrix}
0&0 \\
2\partial _{[m}\ti \lambda_{n]} 
&0 \end{pmatrix}\,
\ee
where
\be
\partial _{[m}\ti \lambda_{n]} 
= -\frac{1}{2}\,
  \partial_m  \xi_{1N}\partial_n \xi_{2}^N
   -(1\leftrightarrow 2)\;
\ee 
 Exponentiation gives the finite $O(D,D)$ matrix $e^\D= 1 +\D$
 \be
\label{eriss}
e^\Delta =  \begin{pmatrix}
1&0 \\
2\partial _{[m}\ti \lambda_{n]} 
&1 \end{pmatrix}\,.
\ee
For any generalised vector $V^M$, the generalised Lie derivative 
$\widehat{\cal L}_{\lambda}$ with parameter $\xi^M= (0, \ti \l _m)$ is
\be
\widehat{\cal L}_{\lambda} V^M = \D ^M{}_N V^N
\ee
with $\D$ given by (\ref{riss}), and
\be
e^{\widehat{\cal L}_{\lambda} }V = e^\D   V
\ee
This extends tensorially to arbitrary generalised tensors.

With this local solution of the strong constraint in a patch $\cal U$ of $M$, the local $O(D,D)$ transformations or $\D$-transformations of \cite{Berman:2014jba} are just the 
DFT gauge transformations   
with parameter $\xi^M= (0, \ti \xi _m)$, acting on the generalised metric through 
antisymmetric tensor gauge transformations with
parameter $\ti \xi_m$.
Then the DFT gauge transformations with parameters $(\xi^m, \ti \xi_m)$, modulo the $\D$-transformations
which are DFT gauge transformations with parameter $ (0, \ti \xi _m)$ are represented by the DFT  gauge transformations with parameter $(\xi^m,0)$. These are just the diffeomorphisms acting on the subspace $U$ with coordinates $x^m$, and can be written in the form in which the coordinates $(x^m, \ti x_m)$ transform as $X\to X'(X)$
\be
x \to x'(x)= e^{-\xi^m \partial _m} x, \qquad \ti x \to \ti x' = \ti x
\ee
The DFT gauge transformations modulo  local $O(D,D)$ transformations are then just the diffeomorphisms of $U$.
Thus in \cite{Berman:2014jba}, the coordinate transformation
 of the doubled space $X\to X'=d(\xi^m, \tilde \xi_m) X$ with
\be
d(\xi^m, \tilde \xi_m)= \exp (- \xi^m \partial _m -\ti \xi _m \ti \partial ^m)
\ee
is mapped to the DFT gauge transformation $h(\xi^m, \tilde \xi_m)$, which modulo antisymmetric tensor gauge transformations is just the diffeomorphism $\exp ( -\xi^m \partial _m )$.
Thus we are obtaining the natural homomorphism from ${\rm Diff} (\cal U)$ to ${\rm Diff} (  U)$ corresponding to $(\xi,\ti \xi)\to (\xi,0)$. However, this loses almost all information about the $\ti \xi $ transformations, and
essentially restricts attention to the subgroup of DFT gauge transformations corresponding to diffeomorphisms of $U$. 
 It would be much more useful to have formulae   for finite gauge transformations for the whole gauge group
with both parameters $\xi$ and $\ti \xi $.

On scalars $\Phi (X)$ satisfying the strong constraint, the diffeomorphism $d(\xi^m, \tilde \xi_m)$ only acts through $\exp ( -\xi^m \partial _m )$, so it is natural  to go from ${\rm Diff} (\cal U)$ to ${\rm Diff} (  U)$.
For generalised tensors, there is also the action of $ {\cal F}$ on the tensor indices, as in 
(\ref{finiteF}).
It will be shown in the next section that in fact
\be
{\cal F} = \hat R e^ \D
\ee
for some $\D$-transformation $\D$, where
\be
\hat R = 
\begin{pmatrix}
\frac{
  \partial x' }
  {\partial x}  &0 \\
0&\frac{
  \partial x }
  {\partial x'} \end{pmatrix}\,
\ee
so that, modulo $\D$-transformations, the action of $\cal F$ is through a simple action of 
${\rm Diff} (U)$. Further, it will be seen that
the composition  of ${\cal F} _1= \hat R _1e^{ \D_1}$ and  ${\cal F} _2= \hat R _2e^{ \D_2}$ is of the form
 \be
 {\cal F} _1{\cal F} _2 = (\hat R _1 \hat R _2 )e^{ \D_{12}}\ee
 for some $\D_{12}$
 so that the composition of the  transformations of \cite{Berman:2014jba} gives the desired result up to 
$\D$-transformations. The gerbe-like properties of the composition rules
for three transformations ${\cal F} _1{\cal F} _2 {\cal F} _3$
 found in \cite{Berman:2014jba} are then seen as a consequence of working modulo 
antisymmetric tensor gauge transformations, given the
role of such  gauge transformations as gerbe transition functions.

In the next section, explicit forms for finite DFT gauge transformations will be found in which the coordinates transform and full information about $\ti \xi $ transformations is kept. They compose with a standard group structure without any non-associativity or gerbe structure. Working with the full transformations rather than modulo the $\D$-transformations will also allow the precise identification of the role of gerbes in the geometry.

\subsection{The Proposal with Coordinate Gauge Symmetry}

In a patch in which the fields and parameters depend on $x^m$ but not $\ti x_m$, the coordinate gauge transformation (\ref{coordgauge}) becomes \cite{Park:2013mpa}
\be
x^m\to x^m, \qquad \ti x_m \to \ti x_m +\ti  \l_m (x)
\ee
where
$\ti \l_m= \phi \partial _m \chi$. 
This is then associated with the DFT gauge transformation with parameter $\r^M =(0,\ti \l _m) $, which as we have seen is an antisymmetric tensor gauge transformation acting through the $O(D,D)$ transformation (\ref{eriss}).
As before, this association of DFT gauge transformations with diffeomorphisms is not a homomorphism.
It was shown in \cite{Park:2013mpa} that the transformation (\ref{finiteF}) with (\ref{Fdef})  with (\ref{xtranssa}) agrees with (\ref{expl}) up to such 
antisymmetric gauge transformations. As in
 \cite{Berman:2014jba}, this gives a form for finite DFT transformations,  but only  up to such
antisymmetric gauge transformations.

\section{Finite Gauge  Transformations}\setcounter{equation}{0}

\subsection{Finite Gauge Transformations for Generalised Vectors }

Consider   a patch $\cal U$ of $  M$ with coordinates $X^M=(x^m,\tilde x_m)$ and  a generalised vector   decomposing as
\be
V^M = \begin{pmatrix}   v^m \\ \tilde v_m \end{pmatrix} \, ,
\ee
  in which 
 the strong constraint is solved by having
 all fields independent of $\tilde x_m$ so that
\be
 \tilde \partial^m  =0
 \ee
 on all fields and parameters, as in section \ref{StrongResolve}.
 Then the fields just depend on the coordinates $x^m$, parameterising a $D$-dimensional patch $U\subset {\cal U}$.
 
The generalised Lie derivative
\be 
 \widehat{{\cal L}}_{V} W^{M} \ = \ V^{P}\partial_{P}W^{M}{}
  +W^{P}\, (\partial^{M}V_{P}
  \, -\partial_{P}V^{M})\,
 \ee
 for $V^M(x),W^M(x)$
 then has the components
 \be 
 (\widehat{{\cal L}}_{V} W)^{m} \ = \ v^{p}\partial_{p}w^{m}{}
    -w^p\partial_{p}v^{m}= {\cal L}_v w^m
 \ee
and
\begin{eqnarray}
(\widehat{{\cal L}}_{V} W)_{m} \   &=& \ v^{p}\partial_{p} \tilde w_{m}{}+ 
  \tilde w_{p}\partial_m v^{p}
  +w^{p} (\partial_{m}\tilde v_{p} 
  \, - \partial_{p}\tilde v_m)\,
  \\ &=& \ 
   {\cal L}_v
      \tilde w  _{m}{}
  +w^{p} (\partial_{m}\tilde v_{p} 
  \, - \partial_{p}\tilde v_m)
\end{eqnarray}
where $ {\cal L}_v  
$  is the usual Lie derivative on $U$.

Under an infinitesimal transformation with parameter $V^M$, suppose $W$ transforms as 
\be
\delta W^M=  \widehat{{\cal L}}_{V} W^{M} 
\ee
giving
\begin{eqnarray}
\delta w^m &=&{\cal L}_v w^m
\\
\delta \tilde w_m &=& {\cal L}_v
      \tilde w  _{m}{}
  +w^{p} (\partial_{m}\tilde v_{p} 
  \, - \partial_{p}\tilde v_m)
\end{eqnarray}
 Now we introduce a gerbe connection\footnote{Here for simplicity we choose the gerbe connection to be the  $b_{mn}$ appearing in the generalised metric. We could choose any other gerbe connection $b'_{mn}$ here,  with $B=b'-b$ a globally defined 2-form, in which case $B$ would appear explicitly in some  of the following formulae, such as the untwisted form of the generalised metric.} $b_{mn}$ on $U$   transforming as
\be
 \delta_{v}b_{mn} = {\cal L}_{v}b_{mn} 
  +\partial_{m}\tilde{v}_{n}-\partial_{n}\tilde{v}_{m}
   \ee
  and define 
  \be
  \hat w_m = \tilde w_m - b_{mn} w^n
  \ee

  Then, remarkably, $ \hat w$ transforms as a 1-form on $U$ under $v$ transformations and is invariant under $\tilde v$ transformations:
  \begin{eqnarray*}
 \delta\hat w_m &=&  {\cal L}_v
      \hat w  _{m}{}
\end{eqnarray*}
 Then given
 \be
W^M = \begin{pmatrix}   w^m \\ \tilde w_m \end{pmatrix} \, ,
\ee
we can define
\be
\hat W^M = \begin{pmatrix}   w^m \\ \hat w_m \end{pmatrix} \, = \begin{pmatrix}   w^m \\ \tilde w_m - b_{mn} w^n
 \end{pmatrix}
\ee
with 
\be
 \delta \hat W^M=  {\cal L}_{v}
 \hat W^M
 \ee
given by the usual Lie derivative on $U$.
$\hat W$ is invariant under $\tilde v$ transformations.

We can then immediately write down the transformations of $w(x,\tilde x)= w(x), \hat w(x,\tilde x)= \hat w(x)$ under finite gauge transformations:
\be
\label{ejgoje}
w'{}^m(x') = w^n(x) \frac {\partial x'{}^m}  {\partial x^n}
\qquad
\hat w'{}_m(x') = \hat w_n(x) \frac {\partial x{}^n}  {\partial x'{}^m}
\ee
where
$x'(x)= e^{-v^m \partial _m} x$.
Moreover, we can use this to find the transformation of $\tilde w$.
The standard global transformations of the gerbe connection can be written as
\be
\label{btrans}
b'{}_{mn}(x') =\Bigl[ b_{pq}(x)
+
(\partial _ p \tilde v_q- \partial _ q \tilde v_p) (x) \Bigr]
 \frac {\partial x{}^p}  {\partial x'{}^m}\frac {\partial x{}^q}  {\partial x'{}^n} 
\ee
This corresponds to doing a $b$-field gauge transformation followed by a 
diffeomorphism (other forms arise by taking these in a different order and give similar results).
We now consider
\be
\hat w'{}_m(x') = \hat w_n(x) \frac {\partial x{}^n}  {\partial x'{}^m}
\ee
We have on the RHS
\be
 \hat w_n(x) \frac {\partial x{}^n}  {\partial x'{}^m} = ( \tilde w_n - b_{np} w^p) \frac {\partial x{}^n}  {\partial x'{}^m}
\ee
while on the LHS
\begin{eqnarray*}
\hat w'{}_m(x') & =&  ( \tilde w'_m - b'_{mn} w'{}^n) (x')
\\ &=& 
\tilde w'_m  (x') - \left[
 b_{pq}(x)
+
(\partial _ p \tilde v_q- \partial _ q \tilde v_p) (x)
\right]
 \frac {\partial x{}^p}  {\partial x'{}^m}\frac {\partial x{}^q}  {\partial x'{}^n}  
w^r(x) \frac {\partial x'{}^n}  {\partial x^r}
\\ &=& 
\tilde w'_m  (x') 
 - \Bigl[
 b_{pq}(x)
+
(\partial _ p \tilde v_q- \partial _ q \tilde v_p) (x)
\Bigr]
w^q(x) \frac {\partial x{}^p}  {\partial x'{}^m}   
\end{eqnarray*}
Putting these together, we find the terms involving $b$ cancel, leaving the transformation for $\ti w$ given by
\be
\label{dfgds}
\tilde w'_m  (x') 
= 
\Bigl[
 \tilde w_n (x)
+
(\partial _ n \tilde v_q- \partial _ q \tilde v_n)w^q (x)
\Bigr]
\frac {\partial x{}^n}  {\partial x'{}^m}
 \ee
 Then (\ref{ejgoje}),(\ref{dfgds}) give the transformation of a generalised vector $W=(w,\ti w)$ under a finite DFT gauge transformation.

Given these forms of finite  gauge transformations, we   now consider their geometric significance.
We have seen that $\hat W =(w,\hat w)$ transforms covariantly under a diffeomorphism of $U$ and is invariant under $\tilde v $ transformations, so that it is a section of $(T\oplus T^*) U$.
On the other hand, $\tilde W =(w, \tilde w)$
is a section of $E$, a deformation of $T\oplus T^*$ resulting from what is sometimes referred  to as twisting
$T\oplus T^*$ by a gerbe.
It is the Courant algebroid defined by the
  short exact sequence \cite{Hitchin:2005in}
$$0 \to T^*  \to E \to T \to 0$$
We will refer to $\hat W$ as the {\it untwisted } form of $ W$, and the transformation $W\to\hat W$
as untwisting a generalised vector.

\subsection{ Generalised Tensors}

The untwisted form $\hat W^M$ of a generalised vector $W^M$
can be written as
\be
\hat W = L W
\ee
where
\be
\label{liss}
L= 
\begin{pmatrix}
1 &0 \\
-b &1  \end{pmatrix}\,
\ee
 denotes the  matrix with components
\be
L^M{}_N = \begin{pmatrix}
\d^m{}_n &0 \\
-b_{mn}
&\d _m {}^n  \end{pmatrix}\,.
\ee
The transformation (\ref{ejgoje}) of the  untwisted vector $\hat W$  is then
\be
\hat W' (X') =\hat R \hat W(X)
\ee
where
\be
\label{hatris}
\hat R = 
\begin{pmatrix}
\L  &0 \\
0& (\L ^{-1})^t  \end{pmatrix}\,
\ee
with
\be
\L ^m {}_n =\frac{
  \partial x'{}^m }
  {\partial x^n}
  \ee
 The coordinate transformation acts only on the $x$:
  \be
X^M \to X'{}^M= \begin{pmatrix} x'{}^m \\   \tilde x'_m \end{pmatrix} \, ,
\ee
with
\be
x^m \to x'{}^m(x), \qquad \ti x_m \to \ti x'_m= \ti x_m
\ee

  The transformation of the twisted vector $W$ was found by twisting the untwisted transformation and is
 \be
 \label{Rtranss}
  W' (X') =  R   W(X)
\ee 
where
\be 
\label{risss}
R= L'(X')^{-1}\hat R L(X) = \hat R S
\ee
and
\be
L'(X')= 
\begin{pmatrix}
1 &0 \\
-b'(x') &1  \end{pmatrix}\,
\ee
with $b'(x')$ given by (\ref{btrans}), and 
\be
\label{radss}
S=  \begin{pmatrix}
\d^m{}_n&0 \\
2\partial _{[m}\ti v_{n]} 
&\d _m {}^n \end{pmatrix}\,.
\ee
The matrices $R,\hat R, L , S$ are all in $O(D,D)$.

Lowering indices with $\eta$ gives similar formulae for a generalised vector with lower index
 \be
U_M = \begin{pmatrix}   \tilde  u_m \\ u^m \end{pmatrix} \, .
\ee
The untwisted vector
\be
\hat U_M =
 \begin{pmatrix}   \hat u_m \\ u^m \end{pmatrix}
 =
  \begin{pmatrix}   \tilde u_m- b_{mn} u^n
 \\ u^m \end{pmatrix}
\ee
transforms with 
\be
 \delta \hat U_M=  {\cal L}_{v}
 \hat U_M
 \ee
and is invariant under $\tilde v$ transformations.
Then the untwisted vector is
\be
\hat U= 
 U L^{-1}
\ee
(i.e. $\hat U_M = U_N (L^{-1})^N {}_M$; recall
$ \eta L \eta ^{-1} = (L^t)^{-1}
$ as $L$ is in $O(D,D)$)
 and transforms under a finite transformation as
 \be
 \hat U'(X') = \hat U(X) \hat R^{-1}
 \ee
For the twisted vectors
 \be
   U'(X') =      U(X) R^{-1}
 \ee

This extends to arbitrary generalised tensors
$T^{MN\dots }{}_{PQ\dots}$.
We define the untwisted tensor
\be 
\hat T^{MN\dots }{}_{PQ\dots} = L^M{}_R L^N{}_S   \dots 
T^{RS\dots }{}_{TU\dots}
  ( L ^{-1})
^T {} _P{}
(   L ^{-1})
^U{}  _Q \dots 
\ee
which transforms as
\be 
\hat T' {} ^{MN\dots }{}_{PQ\dots} (X')= \hat R ^M{}_R  \hat R ^N{}_S   \dots T^{RS\dots }{}_{TU\dots}  (
 \hat R ^{-1})
^T {} _P{}
( \hat R ^{-1})
^U{}  _Q \dots 
\ee
so that the original tensor transforms as
\be 
  T' {} ^{MN\dots }{}_{PQ\dots} (X')=   R ^M{}_R    R ^N{}_S   \dots T^{RS\dots }{}_{TU\dots} 
   ( R ^{-1})
^T {} _P{}
(   R ^{-1})
^U{}  _Q \dots 
\ee
Raising all  lower indices with $\eta$ gives a generalised tensor
$T^{M_1\dots M_p}$ of some rank $p$ which is a section of  $E^p$ while
$\hat T^{M_1\dots M_p}$ is a section of $(T\oplus T^*)^p$.
In particular, 
\be \hat \eta _{MN}= \eta_{MN}
\ee
as $L\in O(D,D)$, and is invariant, $\eta ' = \eta$.

\subsection{The Generalised Metric}

We can now apply the above to the generalised metric.
The untwisted form of the generalised metric 
 \be\label{Hisskoit}
 \hat  {\cal H}_{MN} \ = \    {\cal H}_{PQ} 
   ( L ^{-1})
^P {} _M{}
(   L ^{-1})
^Q{}  _N 
  \ee
 is, using (\ref{His}), simply
\be\label{Hisst}
 \hat  {\cal H}_{MN} \ = \  \begin{pmatrix}    
  g_{mn}   & 
0 \\  0   &
  g^{mn}
   \end{pmatrix}\;
 \ee
and this gives the natural metric on $T\oplus T^*$ arising from $g_{mn}$.
The transformation
 \be\label{Hisskoit}
 \hat  {\cal H}'{}_{MN}(X') \ = \  \hat   {\cal H}_{PQ} (X)
   ( \hat R ^{-1})
^P {} _M{}
(   \hat R  ^{-1})
^Q{}  _N 
  \ee
 simply gives the expected
 \be 
 \label{gtrans}
 g'{}_{mn}(x') = g_{pq}(x) 
  \frac{
  \partial x^p }
  {\partial x'{}^m}
  \frac{
  \partial x^q }
  {\partial x'{}^n}
 \ee
 Finally, the finite  transformation of the (twisted) generalised metric is
 \be\label{Hisskoit}
    {\cal H}'{}_{MN}(X') \ = \      {\cal H}_{PQ} (X)
   (   R ^{-1})
^P {} _M{}
(     R  ^{-1})
^Q{}  _N 
  \ee
  which implies the standard transformations of $g,b$ (\ref{gtrans}),(\ref{btrans}).

\subsection{Large Gauge Transformations}
 
The finite transformations that have been considered above have been obtained by exponentiating   transformations with infinitesimal parameters.
The transformation with parameter $v^m$ exponentiates to give a coordinate transformation under which  
$x \to x'(x)= e^{-v^m \partial _m} x$.
This can then be extended to the symmetry under all coordinate transformations $x \to x'(x)$, not just those obtained from exponentiating infinitesimal diffeomorphisms.
This then gives the group of general coordinate transformations of $U$, which for geometric backgrounds extends to the  group of diffeomorphisms of $N$.

The transformation with parameter $\ti v_m$ exponentiates to give 
 \be
 \label{RtranssS}
  W' (X') =  S   W(X)
\ee 
where $S$ is given by (\ref{radss}), 
using (\ref{Rtranss}),(\ref{risss}) with $\hat R=1$, under which $b \to b +d\ti v$.
This can be extended to replace the exact 2-form $d\ti v$ by any closed 2-form $\omega$ ($d\omega =0$), so that $S$ is now given by 
\be
\label{radssw}
S=  \begin{pmatrix}
\d^m{}_n&0 \\
\omega_{mn}
&\d _m {}^n \end{pmatrix}\,
\ee
so that now $R=\hat R S$ for this $S$ in the formulae or previous sections, and the antisymmetric tensor gauge transformation is $b\to b+\omega$.

The \lq large' gauge transformations are those that are not exponentials of infinitesimal transformations. They consist of large diffeomorphisms, and of $b$-transformations
with $\omega$ closed but not exact.
For a geometric background corresponding to fields on $N$, the gauge symmetry group of DFT is $H={\rm Diff} (N)\ltimes \Lambda ^2_{closed} (N)$, exactly as for the conventional field theory of $g,b,\phi$ on $N$.

\subsection{$\D$-transformations}

In \cite{Berman:2014jba}, it was shown that the
matrix ${\cal F}$ given by (\ref{Fdef}) is given by the matrix $M$ giving the action of finite DFT gauge transformations $e^{\widehat {\cal L}_\xi}$ up to a finite $\D$-transformation
\be
{\cal F}= R e^{\D'}
\ee
for some $\D'$ of the form (\ref{riss}).
Then from (\ref{risss}),
\be
{\cal F}= \hat R e^{\D}
\ee
with $\hat R$ given by (\ref{hatris}) and $e^\D= L e^{\D'}$. Then modulo $\D$-transformations, $\cal F$ is just the matrix $\hat R$ giving the action of the diffeomorphism $x\to x'(x)$ on $T\oplus T^*$.

To find an expression for the product of two $\cal F$'s, we use the fact that  for any matrix of the form
\be
D= 
\begin{pmatrix}
1 &0 \\
B &1  \end{pmatrix}\,
\ee
conjugating with $\hat R$ gives a matrix of the same form:
\be
\hat R ^{-1} D\hat R = D'
\ee
where
\be
D'= 
\begin{pmatrix}
1 &0 \\
B' &1  \end{pmatrix}\, \qquad B'= \L^t B \L
\ee
Then for
\be
{\cal F}_1= \hat R_1 e^{\D _1}, \qquad {\cal F}_2= \hat R_2 e^{\D _2},
\ee
we have
\begin{eqnarray}
{\cal F}_1{\cal F}_2 &=& \hat R_1  \hat R_2  ( \hat R_2 ^{-1}e^{\D _1} \hat R_2) e^{\D _2}
\nonumber
\\
&=&
\nonumber
\hat R_1  \hat R_2  e^{\D '_{12}}
\\
&=&
  R_1    R_2  e^{\D _{12}}
  \label{dgdssdg}
\end{eqnarray}
where $\D _{12}, \D ' _{12}$ are matrices of the form (\ref{riss}).
Then  the result of  the composition of the $\cal F$ matrices agrees with the composition of finite DFT gauge transformations up to a $\D$-transformation, as argued in  \cite{Berman:2014jba}.

In \cite{Berman:2014jba}, double geometries were considered in which fields in  patches were related by transition functions that are  DFT gauge transformations. These transition functions can be given by a coordinate transformation and an action of the matrix $R$ on tensor indices, as we have seen.
In a triple overlap of patches, the three transition functions $R_1,R_2,R_3$ in the three double overlaps must
satisfy
 $R_1    R_2R_3=1$ for consistency.
 However, writing the transformations in terms of $\cal F$ instead of $R$  gives a product  ${\cal F}_1{\cal F}_2 {\cal F}_3$ 
 which is 
 not $1$ but  gives a $\D$-transformation.
 In \cite{Berman:2014jba},  it is suggested that this reflects a gerbe structure of the double geometry.
 Here we see that this is a consequence of writing the $R$ transformations as 
 $\cal F$  transformations up to $\D$ transformations,
 and working only modulo  $\D$ transformations.
 By a similar argument to that leading to (\ref{dgdssdg}), for any $R_1,R_2,R_3$
\be
{\cal F}_1{\cal F}_2 {\cal F}_3
=  R_1    R_2R_3  e^{\D _{123}}
\ee
for a 
${\D _{123}}$ of the form (\ref{riss}).
In particular, if $R_1    R_2R_3=1$, then in general
${\cal F}_1{\cal F}_2 {\cal F}_3$ is not $1 $ but gives a $\D$-transformation. Such issues with the composition of transformations are avoided by using the form of the transformations using the $R$-matrices instead of the one involving the ${\cal F}$ matrices.

\section{Discussion}
 
For DFT in a local patch $\cal U$ with constant $\eta$, the strong constraint leads to fields depending on the coordinates $x$ of a $D$-dimensional subspace $U\subset {\cal U}$,  and  independent of the remaining coordinates $\tilde x$. Simple finite transformations have been found for the
DFT gauge symmetries, and these  encode the gauge symmetries of the   underlying field theory.
These then give the transformations and transition functions for generalised tensor fields.
In the case of a geometric background, the patches $U$ cover a manifold $N$, while $g$ and $H=db$ are well-defined tensor fields on $N$. We will first discuss generalised tensors for this geometric case, and then briefly consider the more general case.

A geometric background consists of a space-time $N$ with fields $g(x),b(x),\phi(x)$ depending on the coordinates $x^m$ of $N$.
In the DFT formulation, $N$ is    a submanifold of a doubled manifold $M$
with coordinates $X^M=(x^m,\ti x_m)$, and the fields are independent of the extra coordinates $\ti x_m$ of $M$.
The DFT gauge transformations have been seen to correspond to a diffeomorphism of $N$ in which
$x\to x'(x)$,  together with a $b$-field gauge transformation with finite parameter $\ti v_m$,
 so that the DFT gauge group is $H={\rm Diff} (N)\ltimes \Lambda ^2_{closed} (N)$.

There are three distinct kinds of \lq vector field' on $M$, all of which have components that can be written as $W^M(X)$, 
but which are sections of different bundles and so transform differently.
First, a conventional vector field on $M$ is a section of the tangent bundle $TM$ of $M$.
It transforms under diffeomorphisms of $M$
as
\be
W'(X') ^M = \frac {\partial X'{} ^M }  {\partial X^ N}\,  W(X) ^N 
\ee
 for any coordinate transformation $X\to X'(X)$
 and such transformations provide the 
 transition functions between patches.
Next, there are generalised vector fields and untwisted generalised vector fields which satisfy the strong constraint by being independent of $\ti x$, so they are fields on $N$.
The
untwisted generalised vector fields $\hat W^M(x)$
are sections of the generalised tangent bundle $T\oplus T^*$ of $N$, and transform under diffeomorphisms   $x\to x'(x)$  of $N$ as
\be
\hat W' (x') =\hat R \hat W(x)
\ee
where
\be
\hat R = 
\begin{pmatrix}
\L  &0 \\
0&(\L ^{-1})^t  \end{pmatrix}\, \qquad \L ^m {}_n =\frac{
  \partial x'{}^m }
  {\partial x^n}
\ee
The  transition functions for such vectors between patches of $N$ is through  this action of the diffeomorphisms.
Finally,  generalised vector fields $ W^M(x)$ 
are sections of $E(N)$, which is  the generalised tangent bundle $T\oplus T^*$ of $N$,
twisted by a gerbe. They transform under the DFT gauge group $H={\rm Diff} (N)\ltimes \Lambda ^2_{exact} (N)$
as
\be
  W' (X') =  \hat R \, S \,  W(X)
\ee 
where
\be
 S=  \begin{pmatrix}
\d^m{}_n&0 \\
2\partial _{[m}\ti v_{n]} 
&\d _m {}^n \end{pmatrix}
\ee
These are patched together by DFT gauge transformation transition functions.

Similarly,
given a field with components $T^{MN\dots P}(X)$, it is necessary to specify whether it is a tensor, a generalised tensor or an untwisted generalised tensor.
In DFT, a key role is played by the constant matrix
\be
\eta_{MN} =  \begin{pmatrix}
0&1 \\1&0 \end{pmatrix}
\ee
If these were the components of a tensor on $M$, i.e. if $\eta $ were a section of $(T^*\otimes T^*)M$, then the presence of a flat metric on $M$ would be highly restrictive and imply that $M$ is locally a flat space.
Moreover, under a change of coordinates $X\to X'(X)$ of $M$, $\eta$ would transform to a new matrix of components that would no longer be constant in general.

If, however, the constant matrix $\eta$  gives the components of a generalised tensor in $(E^*\otimes E ^*)N$, then this places no restriction on $N$ or $M$, and $\eta $ is in fact invariant under 
$H={\rm Diff} (N)\ltimes \Lambda ^2_{closed} (N)$.
Untwisting gives the same matrix as $\hat \eta =  \eta$, now regarded as   a section of
$[(T\oplus T^*)\otimes (T\oplus T^*)] N$. $\eta $ is the natural metric on $(T\oplus T^*)N$
and is invariant under ${\rm Diff} (N)$.
In DFT, there is an $\eta$  which is a generalised tensor and is defined in this way for any manifold $M$ with submanifold $N$.

The constraint $\eta ^{MN} \partial _M\partial _N=0$
is imposed  locally in  patches in DFT, as has been done here. If
    $\eta ^{MN} $ were the components of a tensor, this condition can be extended to a globally well-defined condition.
    However,
   if  $\eta$ is not a tensor but a generalised tensor, then 
   there are problems in extending this form of the  constraint globally.
   In the case of a geometric background, then one can simply use the form of the constraint $\ti \partial^m=0$, so that the fields are fields on $N$.
   
Then DFT formulates a conventional field theory on $N$ in terms of generalised geometry, based on the generalised tangent bundle 
$(T\oplus T^*)N$. Type II supergravity has been formulated in terms of generalised geometry  in \cite{Coimbra:2011nw}.

More generally, there may not be a geometric background $N$, and the above need apply only locally.
The doubled manifold $M$ is   covered by  patches $\cal U$  with coordinates  $(x,\ti x)$ in each of which
there is a \lq physical' subspace $U$ with coordinates $x$, and the DFT then gives a field theory on each $U$ formulated in terms of generalised geometry.
However, the patches $U$ may not fit together to form a submanifold $N$ in general, 
and may instead give a T-fold. 
 If the patches are glued together only with transition functions that are DFT gauge symmetries 
$x\to x'(x)$, they can form a manifold $N$, but if $O(D,D)$ transformations are also involved, then a non-geometric space can result. Generalised vectors are then defined over $U$ as sections of the bundles
$(T\oplus T^*)U$ or $E(U)$ over $U$.
Transition functions and non-geometric spaces in DFT will be discussed in a separate paper.

\section*{Acknowledgments}  
I would like to thank Dan Waldram for his collaboration in early stages of this work, and for invaluable discussions.
I would also like to thank O~Hohm, D~Lust and
B.~Zwiebach for useful discussions.
This work was supported  by STFC grant ST/L00044X/1 and EPSRC grant EP/K034456/1.

\end{document}